 
 
\parindent 0pt
\magnification\magstep1
\baselineskip 16pt plus 1pt minus 0.5pt
\hsize = 15.0 true cm
\vsize = 24.0 true cm
\abovedisplayskip 16pt plus 5pt minus 6pt
\belowdisplayskip 16pt plus 5pt minus 7pt
\setbox\strutbox = \hbox{\vrule height 13.5 true pt depth
                          6.5 true pt width 0pt} 

\hfill  22 June 1992
 
\vskip2.0truecm

\centerline{\bf Response Function of Asymmetric Nuclear Matter} 

\vskip1.5truecm
\centerline{Kazuo Takayanagi}
\centerline{\it Department of Mathematical Sciences, Tokyo Denki University} 
\centerline{\it Hatoyama, Saitama 350-03, Japan}
\vskip0.8truecm
\centerline{and } 
\vskip0.8truecm
\centerline{Taksu Cheon }
\centerline{\it Department of Physics, Hosei University} 
\centerline{\it Chiyoda-ku Fujimi, Tokyo 102, Japan} 
\vskip2.5truecm

\centerline{\bf Abstract}
\vskip1truecm
The charge longitudinal response function is examined in the framework 
of the random-phase approximation in an isospin-asymmetric nuclear matter 
where proton and neutron densities are different.  
This asymmetry changes the response through both the particle-hole interaction 
and the free particle-hole polarization propagator.  
We discuss these two effects on the response function 
on the basis of our numerical 
results in detail.
\vfill\eject

\qquad 
Since the development of experimental devices has enabled 
us to separate 
the charge longitudinal (Coulomb) response of nuclei 
from the spin transverse 
one in the electron scattering [1,2,3],
these two response functions have been 
widely investigated both experimentally and theoretically [4].  
Empirical response functions are now available for a variety
of medium and heavy nuclei 
ranging from 40Ca to 208Pb, most of which have neutron excess.   
In order to examine the role of the neutron excess, 
several theoretical attempts 
have been made to calculate the response function of isospin-asymmetric 
nuclear matter taking into account the different phase space available for 
proton and neutron ph states [5,6,7].  This should not be, however, 
the whole story, because the asymmetry changes also the effective 
ph interaction [8].  
In this letter, we examine how these two effects modify the charge 
longitudinal response 
function of the isospin-asymmetric nuclear matter, i.e. 
(i) the different ph phase space for protons and neutrons, and 
(ii) the change of the effective ph interaction.

\qquad 
We consider a nuclear matter of which the proton and the neutron 
densities are  
given by $\rho_p$, $\rho_n$ respectively.  Let the $\varepsilon$ stand for 
the size of the asymmetry:

$$\varepsilon \ =\ {{\rho _p-\rho _n} \over {\rho _p+\rho _n}}
\quad. \eqno(1)$$

First, let us see how this asymmetry enters the expression for the response
function $R({\vec q}, \omega)$ in 
the random-phase approximation (RPA) [6, 7], 
which is given by

$$R({\vec q},\omega )=O^+\ {{\Pi ({\vec q},\omega )} 
\over {1-V\ \Pi ({\vec q},\omega )}}\ O 
\quad. \eqno(2)$$

where $\vec q$ and $\omega$ are the momentum and the energy transfers 
by the external operator $O$.  
The symbols $\Pi$ and $V$ are the free ph propagator (Lindhard function) and 
the effective interaction ($g$-matrix), respectively.  
In the isospin-asymmetric nuclear matter, $\Pi$ and $V$ deviate from
their symmetric nuclear matter values, 
and therefore can be expanded in powers of $\varepsilon$ as

$$V\ =\ V^0+V^1+V^2+\ \cdots  
\quad. \eqno(3)$$

$$\Pi \ =\ \Pi ^0+\Pi ^1+\Pi ^2+\ \cdots 
\quad. \eqno(4)$$

where the indices represent the order in powers of $\varepsilon$.
Substituting eqs. (3) and (4) into eq. (2), 
we see that in order to obtain the response exactly up to some order in 
$\varepsilon$, 
we have to calculate both $\Pi$ and $V$ up to the same order. 

\qquad
We proceed in the following way.  
First, we explain the effective interaction, and then  examine 
the response function.

\qquad
The $g$-matrix in an isospin-asymmetric nuclear matter satisfies 
the Bethe-Goldstone equation formally identical to the one in 
the symmetric nuclear matter, i.e.

$$g\ =\ v\ +\ v\ G\ Q\ \ g
\quad. \eqno(5)
$$

where v is the bare two-nucleon interaction (Reid soft core potential), $G$,
 the two nucleon propagator including the self-energy correction,
and $Q$, the Pauli exclusion operator.  
The difference of proton and neutron densities $\rho_p$ and $\rho_n$ generates 
the isovector component in Q in addition to the isoscalar component 
present in the symmetric nuclear matter [8].  
A detailed description of the resultant interaction will be discussed 
in a forthcoming paper [9].  Here we just show the part of the interaction 
which is relevant to the charge longitudinal response in the  framework of RPA:

$$g\ =\ g^0\ +\ g^\tau \ \tau _1\cdot \tau _2\ 
+\ g^\alpha \ \ (\tau _1+\tau _2)\ +\ \cdots 
\quad. \eqno(6)
$$

where $\tau _1$ and $\tau _2$ are the isospin operators 
of two interacting nucleons.  
In the above expression, $g^0$  and $g^\tau$ already exist in the $g$-matrix 
in the symmetric nuclear matter, and  $g^\alpha$ is the new component 
which appears due to the isospin asymmetry 
and is of first order in $\varepsilon$.  
The $g$-matrix of eq. (6) is invariant under the exchange of 
all protons and neutrons including the medium, 
because $\varepsilon$ changes its sign under such transformation.  
The ph interaction is derived from the above $g$-matrix as

$$V_{pp}\ =\ g^0+g^\tau +2g^\alpha  \quad, $$
$$V_{nn}\ =\ g^0+g^\tau -2g^\alpha  \quad, \eqno(7)$$
$$V_{pn}\ =\ V_{np}\ =\ g^0-g^\tau  \quad $$

where $V_{pp}$, for example, is the interaction 
which describes the transition from a proton ph to another proton ph.  
We show  the real and the imaginary parts of $g^0$, $g^\tau$ and 
$g^\alpha/\varepsilon$
in figures 1-a  and 1-b, respectively, as functions of the energy.  
Note that the imaginary part of the effective interaction is 
the realization of the two-particle emission in the multiple 
scattering process described by the Bethe-Goldstone equation, eq. (5).  
We can see in the figure that the effect of the asymmetry appears 
mainly in the imaginary part of , and increases in the low energy side.  
This is because the asymmetry is more important there 
than in the high energy region.  

\qquad
Let us now turn to the response function.  
It is straightforward to write down the expression for 
the Lindhard function in the isospin asymmetric nuclear matter.  
Let $\Pi ^p$  and $\Pi ^n$  be the Lindhard functions for 
proton and neutron, respectively.  
The total polarization propagator of eq. (4)  can be written as

$$
\Pi \ =\ {{1+\tau _z} \over 2}\Pi ^p+{{1-\tau _z} \over 2}\Pi ^n
\quad, \eqno(8)
$$

\qquad
Actual calculations of the response $R$ of eq.(2) are carried on for
the charge longitudinal operator

$$
O\ =\ {{1+\tau _z} \over 2}\ exp(i\ {\vec q}\cdot {\vec r})
\quad, \eqno(9)
$$

where we take the momentum transfer q = 2.0 fm$^{-1}$.    
Substituting equations (8) and (9) into eq. (2), 
we arrive at the following expression [6] for 
the response function $R({\vec q}, \omega)$:

$$
R({\vec q},\omega )\ =\ {{\Pi ^p({\vec q},\omega )} 
\over {1-\tilde V_{pp}\ \Pi ^p({\vec q},\omega )}}
\quad, \eqno(10)
$$

where $\omega$ is the energy transfer from the external field, 
and $\tilde V_{pp}$ is a renormalized proton ph interaction which takes 
into account the neutron ph states, and is given by

$$
\tilde V_{pp}\ =\ V_{pp}+V_{pn}{{\Pi ^n} 
\over {1-V_{nn}\Pi ^n}}V_{np}
\quad, \eqno(11)
$$

The strength function, which is the observable 
in the quasielastic scattering, is then given by

$$-{1 \over \pi }Im\ R\ =\ 
{1 \over {\ \left| {\ 1-\tilde V_{pp}\Pi ^p\ } \right|^2}}
\left\{ {Im\ \Pi ^p+\left| {\ \Pi ^p\ } \right|^2Im\tilde V_{pp}} 
\right\}
\quad, \eqno(12)
$$

\qquad  
Now we present our numerical results. 
In the actual calculation, we have taken the proton and 
neutron fermi momenta as 
$k_F^p$ = 1.20 fm$^{-1}$ and $k_F^n$  = 1.39 fm$^{-1}$, 
which correspond to $\varepsilon$ = $-0.213$ in eq. (1) 
(note that  $\varepsilon$ = $-0.2$ on average in ${}^{48}$Ca).  
Our results of the response function are shown in fig. 2.  
The dotted curve represents the free response ($-1/\pi \ Im \Pi ^p$), and 
the dashed curve stands for the response of the symmetric nuclear matter,
$R_0$, which is obtained by setting $\varepsilon$ = 0 in eq. (12).  
The solid curve is the response function in the asymmetric 
nuclear matter, eq. (12). 
The dotted-dashed curve shows the same response, 
but with the real part of the ph interaction alone.  
The difference between the dotted and the dashed curve shows 
the well known feature of the charge longitudinal response: 
the enhancement in the low energy side due to the attractive 
isoscalar interaction, and the quenching in the high energy side due to 
the isovector repulsive interaction [10].  
A comparison of the dashed and the full curve immediately shows 
that the effect of the asymmetry slightly enhances the response  
in the quasielastic region, and quenches in the lower and higher energy side, 
as compared to the response of the symmetric nuclear matter.  
In the following, we look into our numerical results more closely.

\qquad 
First, we examine the effect of 
the imaginary part of our ph interaction, i.e., 
the difference between the dotted-dashed and the solid curve.  
We observe that the role of the imaginary part of the ph interaction is 
to remove the transition strength from the low energy side to the quasielastic 
and higher energy region where two-nucleon emission channel 
is more widely open.  
Second, in order to see separately how the asymmetry 
in $V$ and in $\Pi$ affect the response, 
we define two response functions $R_V$ and $R_{\Pi}$
which are respectively obtained by taking 
into account the change of the interaction and the ph propagator alone.  
In fig. 3, we show the differences of $R_V$ and $R_{\Pi}$ from 
the response of 
symmetric nuclear matter ($R_0$) by the dotted and the dashed curve.  
We also show the difference of the full response, 
$-1/\pi \ Im(R-R_0)$, by the solid curve.  
The figure shows that the asymmetry in the interaction lowers 
the response of the symmetric nuclear matter, while the asymmetry in the 
Lindhard function gives a sizable enhancement as a whole.  
Next, we compare the whole strength integrated 
over all the energy transfer, i.e., 
the Coulomb sum rule [4, 11, 12].  
We find that $R_V$ and $R_{\Pi}$ yield a 6 $\%$ reduction and 
a 3 $\%$ enhancement of the sum rule, 
and $R$ a 3 $\%$ reduction, compared to the response of 
the symmetric nuclear matter, .

\qquad
We now focus on the energy dependence of the three curves in the figure.  
The behavior of $-1/\pi \ Im(R_V-R_0)$, the dotted curve, 
can be understood as follows:  
first, the asymmetry here ($\rho _p < \rho _n$)
enlarges the two-proton emission channel 
compared to the symmetric case ($\rho _p = \rho _n$), 
because the Pauli blocking is less operative for protons - thereby 
the imaginary part of $V_{pp}$ of eq.(7) is negative and 
larger in magnitude than its symmetric limit  $g^0 + g^{\tau}$ .  
Looking at this effect (enlargement of the two-proton emission channel) 
as a function of the ph energy $\varepsilon$, 
we recognize that (i) for a large $\omega$, 
this effect is fully operative because the whole decrement of the phase 
volume excluded by the Pauli blocking turns into the increment of 
the phase space 
available for emitted protons, while 
(ii) for a small w, this effect is less operative 
because only a part of the above decrement is available.  
The combined effect on the response function is to 
remove the transition strength
 from a low energy side to a high energy (several hundred MeV) region 
where more two-proton emission channels are coupled to the response, 
therefore the behavior of the dotted curve.

\qquad
The asymmetry in the Lindhard function affects 
the response in a subtler way.  
Here we just explain why this causes a larger deviation from 
the symmetric case in the high energy region 
than the asymmetry in the ph interaction.  
In fig. 4, a transition of a ph state to another is shown diagrammatically.  
The intermediate states of the multiple scattering 
process (A in the figure) is  
accounted for by the $g$-matrix, 
and does not appear explicitly in the actual calculation.  
On the other hand, the ph phase space (B in the figure)  
is taken into account by 
the Lindhard function explicitly.  
Now let us look at the asymmetric part of 
the proton and neutron phase space in spaces A and B.  
Because of the short range nature of the bare nucleon-nucleon interaction, 
the phase space A is located much higher in energy (several hundred MeV) 
than B (around hundred MeV), which makes the asymmetric part of 
the phase space is less important in A than in B.  
This is the reason why the asymmetry plays 
a more important role in $R_{\Pi}$
than in $R_V$ in the high energy (around one hundred MeV) region.

\qquad
In summary, we have pointed out that in order to calculate a response function 
of asymmetric nuclear matter, it is necessary to account for 
the effect of the asymmetry both 
in the effective interaction and in the ph propagator.  
We have shown, on the basis of our numerical results, 
that these two effects are opposite in sign and comparative in magnitude.  
We have also explained how and why these two effects appear.  
It is now clear that further study of the response function of N $\ne$ Z 
nuclei should simultaneously take these two effects into account.

\vskip1.0truecm

\indent 
The numerical work has been performed on SPARCStation 2 
at Tokyo Denki University and at Department of Physics, Hosei University, 
and also on VAX6000/440 of Meson Science Laboratory, University of Tokyo.  
This work has been financially supported by the Research Institute for 
Technology at Tokyo Denki University.

\vfill\eject

{\bf References}
\vskip1.0truecm

 \item{1.}   {P. Barreau et al., Nucl. Phys. {\bf A402}, 515 (1985).}

 \item{2.}   {Z. E. Meziani et al.,  Phys. Rev. Lett. {\bf 52}, 2130 (1984);
  {\bf 54}, 1233 (1977).}

 \item{3.}   {C. C. Blatchley et al., Phys. Rev. {\bf C34}, 1243 (1986).}

 \item{4.}   {B. Frois and C. N. Papanicolas, 
  Ann. Rev.Nucl. Part. Sci. {\bf 37}, 133 (1987).}

 \item{5.}   {S. Stringari and E. Lipparini, Nucl. Phys. {\bf A473}, 61 (1987).}

 \item{6.}   {W. M. Alberico, G. Chanfray, M. Ericson and A. Molinari, 
  Nucl. Phys. {\bf A475}, 233 (1987).}

 \item{7.}   {W. M. Alberico, A. Drago and C. Villavecchia, 
  Nucl. Phys. {\bf A505}, 309 (1989).}

 \item{8.}   {T. Cheon and K. Takayanagi, 
 Phys. Rev. Lett. {\bf 68}, 1292 (1992).}

 \item{9.}   {T. Cheon and K. Takayanagi, in preparation.}

 \item{10.}   {Shigehara, K. Shimizu and A. Arima, Nucl. Phys., 
 {\bf A492}, 388 (1989).}

 \item{11.}   {K. Takayanagi, Nucl. Phys. {\bf A516}, 276 (1990);
  {\bf A522}, 494 (1991); {\bf A522}, 523 (1991).}

 \item{12.}   {G. Orlandini and M. Traini, 
 Rep. Prog. Phys. {\bf 54}, 257 (1991).}

\vfill\eject

{\bf Figure Captions}

\vskip1.0truecm
Fig. [I]:  The g-matrix of eq. (6) as a function of the starting energy Ep.  
(a) The real part, and (b) the imaginary part.

\vskip0.5truecm
Fig. [II]: The charge longitudinal response of nuclear matter.  
The dotted and the dashed curve show the free  response and 
the response of the symmetric nuclear matter, respectively.  
The solid curve shows the response of the asymmetric nuclear matter of eq. (12).  
The dotted-dashed curve is the same as the solid curve, 
but with the real part of the ph interaction alone.

\vskip0.5truecm
Fig. [III]: The changes of the free response when the modification of 
the interaction alone  (dotted line), of the ph propagator alone (dashed line) 
and both are taken into account (solid line).

\vskip0.5truecm
Fig. [IV]: A transition process of a ph state into another.  
See the text.

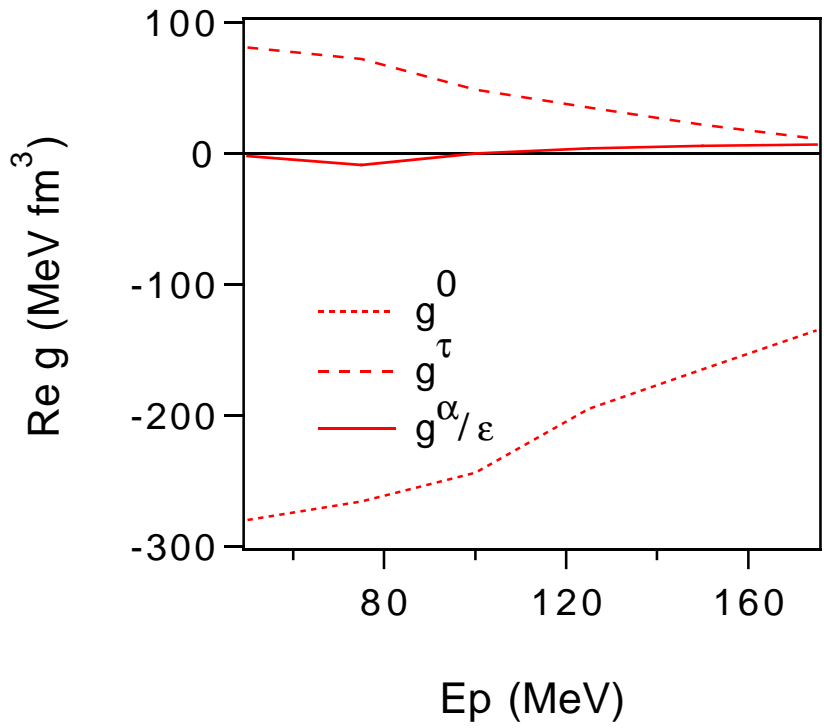
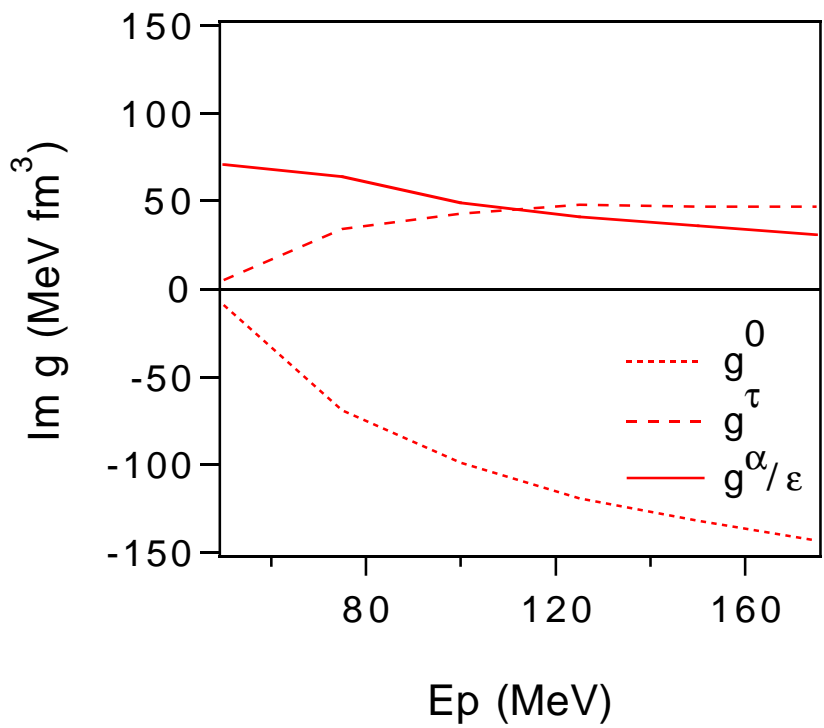
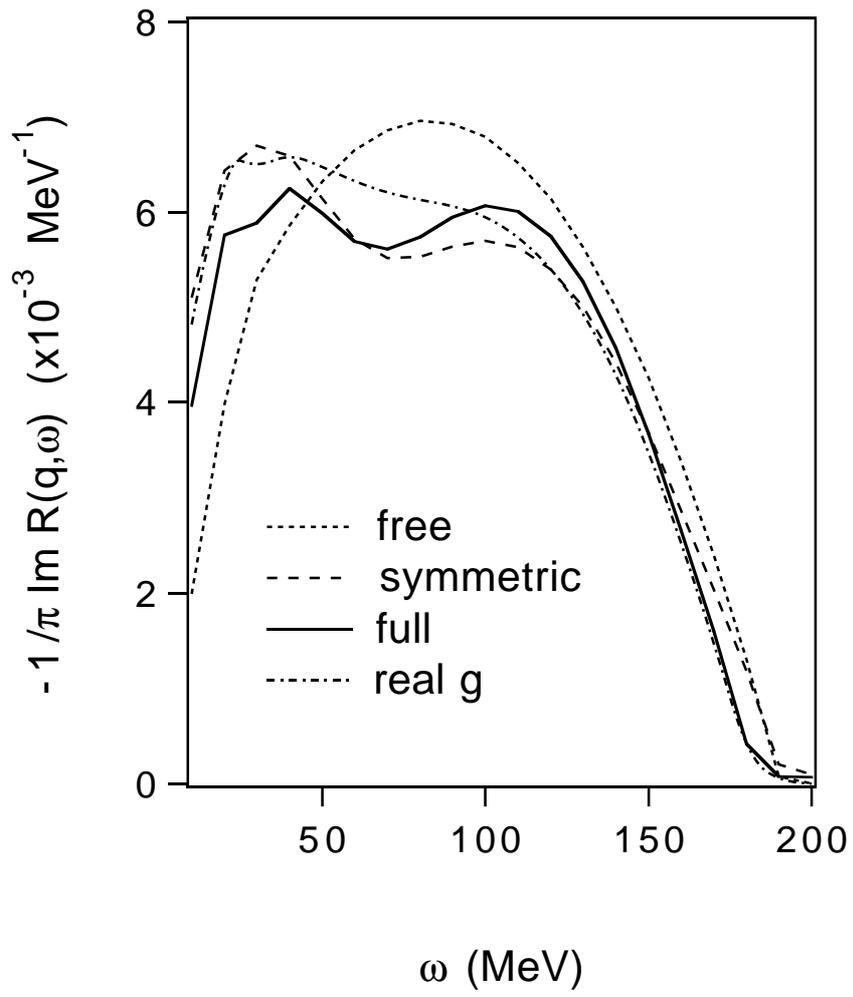
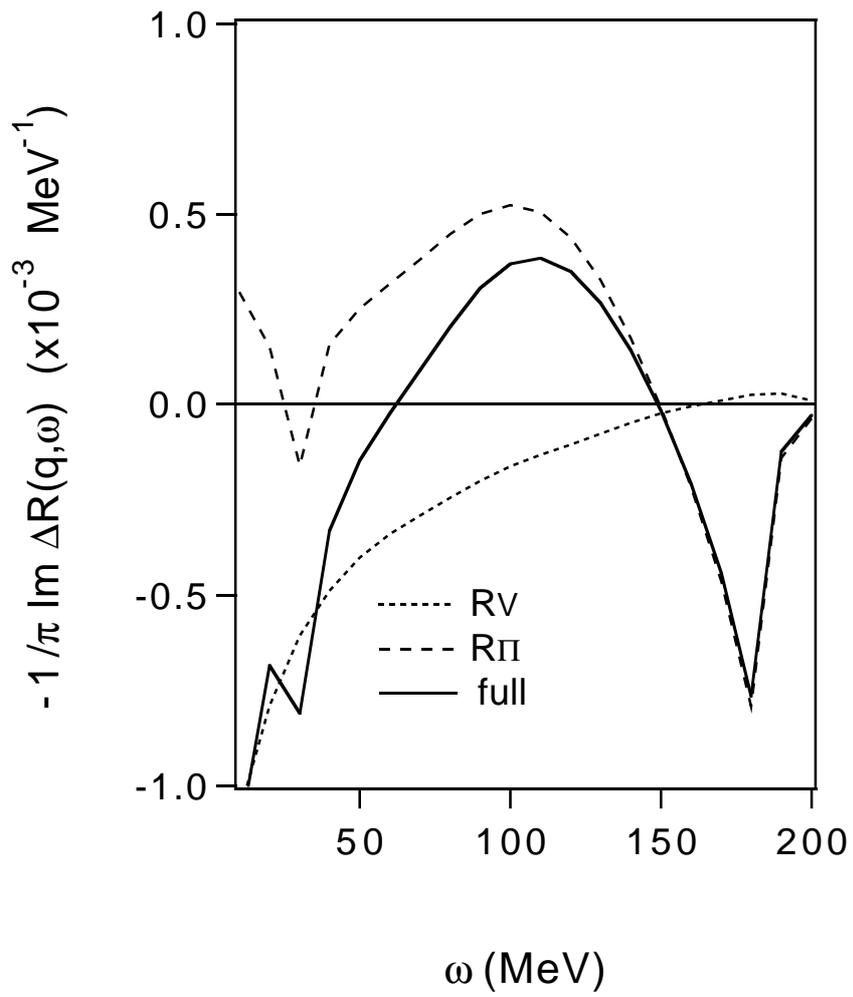
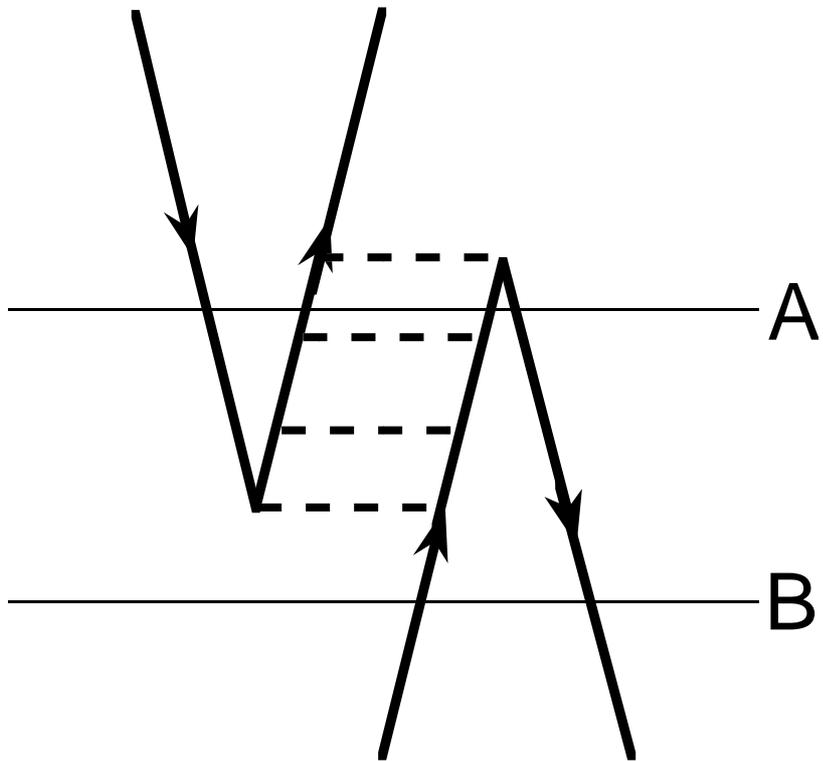
\bye